# Material Effects on Electron Capture Decays in Cryogenic Sensors

Amit Samanta[1], Stephan Friedrich[1], Kyle G. Leach[2], Vincenzo Lordi[1,*]

[1] Lawrence Livermore National Laboratory, 7000 East Ave., Livermore, CA 94550, USA
[2] Colorado School of Mines, 1500 Illinois St., Golden, CO 80401, USA

**Abstract**

Several current searches for physics beyond the standard model are based on measuring the electron capture (EC) decay of radionuclides implanted into cryogenic high-resolution sensors. The sensitivity of these experiments has already reached the level where systematic effects related to atomic-state energy changes from the host material are a limiting factor. One example is a neutrino mass study based on the nuclear EC decay of $^7$Be to $^7$Li inside cryogenic Ta-based sensors. To understand the material effects at the required level we have used density functional theory and modeled the electronic structure of lithium atoms in different atomic environments of the polycrystalline Ta absorber film. The calculations reveal that the Li 1$s$ binding energies can vary by more than 2 eV due to insertion at different lattice sites, at grain boundaries, in disordered Ta, and in the vicinity of various impurities. However, the total range of Li 1$s$ shifts does not exceed 4 eV, even for extreme amorphous disorder. Further, when investigating the effects on the Li 2$s$ levels, we find broadening of more than 5 eV due to hybridization with the Ta band structure. Materials effects are shown to contribute significantly to peak broadening in Ta-based sensors that are used to search for physics beyond the standard model in the EC decay of $^7$Be, but they do not explain the full extent of observed broadening. Understanding these in-medium effects will be required for current- and future-generation experiments that observe low-energy radiation from the EC decay of implanted isotopes to evaluate potential limitations on the measurement sensitivity.

## 1. Introduction

Some of the most accurate measurements of nuclear decays can be made by embedding radionuclides of interest in cryogenic high-resolution radiation sensors. These state-of-the-art detection methods are capable of eV-scale resolution and allow measuring the total energy of a nuclear decay, including the daughter recoil and atomic de-excitations, with unprecedented precision. This approach is used to great effect in the current generation of searches for physics beyond the standard model (BSM) in the neutrino sector [1 – 3]. In these experiments, ultra-high sensitivity measurements of the low-energy nuclear decay products are performed to reconstruct the missing momenta that would result from heavy neutrino mass states.

Three experiments of this type are currently active that use the nuclear electron capture (EC) decay of $^{163}$Ho or $^7$Be radioisotopes embedded in either magnetic microcalorimeters (MMCs) [1], transition edge sensors (TESs) [2], or superconducting tunnel junctions (STJs) [3]. The high energy resolution of these cryogenic detectors provides the required sensitivity in the search for BSM physics. This feature also makes these experiments sensitive to the chemical shifts due to interactions between the radioisotope's atomic shell and the matrix material into which it is embedded. Effects of this nature are so small that they have traditionally been neglected in decay experiments due to the large difference in energy scales for nuclear processes and chemical shifts. In this paper, we show that this is no longer possible for experiments sensitive to energies and masses on the eV-scale or below where these effects become significant– specifically in the case of the $^7$Be decay in Ta-based STJ detectors.

---

[*] Corresponding author email: lordi2@llnl.gov

The <u>Be</u>ryllium-7 <u>E</u>lectron capture in <u>S</u>uperconducting <u>T</u>unnel junctions (BeEST) experiment is a search for BSM physics in the neutrino sector using $^7$Be implanted into Ta-based superconducting tunnel junctions (STJs) [3, 4]. When $^7$Be decays by EC to $^7$Li and an electron neutrino inside of the thin-film STJ, the neutrino escapes, and the kinetic energy of the recoiling $^7$Li daughter atom can be measured with high precision. The core hole produced by the capture of the bound electron relaxes on a short enough time scale so that its energy adds to the $^7$Li reoil signal within the timing resolution of the sensor. The resulting low-energy decay spectrum therefore consists of four peaks due to EC from the $^7$Be 1$s$ shell (K-capture) or the $^7$Be 2$s$ shell (L-capture) into the ground state or the first exited state of $^7$Li (Fig. 1). Any admixture of a heavy BSM neutrino mass state, or in fact any new massive particle that couples to the decay, would reduce the recoil energy and add a shifted $^7$Li spectrum at lower energies as a signature.

The absorber films in the STJ sensors of the BeEST experiment consist of polycrystalline Ta in the body-centered cubic (bcc) phase deposited by sputtering [4]. Radioactive $^7$Be nuclei ($T_{1/2}$ ~ 53 days) are produced via the isotope separation online (ISOL) technique [5] and implanted at an energy of 24 keV through a Si collimator to an average depth of ~55 nm [3, 4]. The ISOL production process also introduces a stable $^7$Li contaminant into the beam with roughly ~50 times higher intensity relative to the $^7$Be, which is simultaneously implanted into the Ta with the same depth profile.

Matrix-dependent effects are suggested by the observation that the EC decay peaks in the recoil spectrum were broadened well beyond the intrinsic energy resolution of the STJ (Fig. 1, black) [4]. The K-shell capture peak at 107 eV had a width of 6.7 eV FWHM, and the L-shell capture peak at 57 eV was even broader with a width of 8.2 eV FWHM. In contrast, the pulsed laser calibration spectrum (Fig. 1, grey), which consists of a comb of peaks due to integer number of absorbed laser photons, shows the expected detector resolution of ~2 eV FWHM in the same energy range. Both spectra were taken with the same STJ sensor at the same time under identical conditions, with the calibration recorded in coincidence with the laser trigger and the EC spectrum in anticoincidence. This broadening of the EC spectrum beyond the calibration signal is critically significant for the BeEST experiment, because it affects the measurable spectral shifts and thus currently limits the sensitivity in the search for neutrinos with masses below 100 keV/c$^2$ where their relevance as dark matter candidates is highest [6].

Although the source of the full broadening effect is unknown, it could be caused by variations of the core hole energies that are added to the recoil (Fig. 1, inset). This is the subject of this work. Generally, for atoms embedded in a material, the core hole energies can change either due to structural differences of the local sites in the host matrix, hybridization of defect states with the orbitals of the matrix materials, or due to impurities that affect the local crystal field. A systematic analysis of the relative contributions from these three effects is needed to understand the sources of peak broadening and to improve the design of the detectors and the experiment. For a direct comparison with recent observations [3, 4], the first system we have chosen to study is that of $^7$Be EC decay in tantalum.

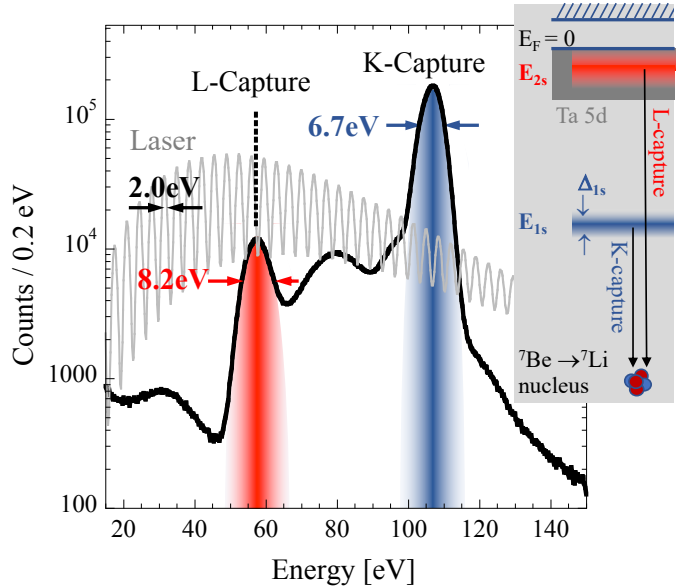

Figure 1: EC spectrum of $^7$Be implanted into Ta-based STJ detectors [3]. The energies from the relaxation of the 1s (blue) or 2s (red) core hole add to the $^7$Li recoil energy. The inset illustrates how variations $\Delta_{1s}$ and $\Delta_{2s}$ in the $^7$Li 1s and 2s binding energies can broaden the EC peaks beyond the resolution of the laser calibration spectrum (grey).

This paper examines the range of $^7$Li 1$s$ and 2$s$ binding energies at different sites in a polycrystalline Ta matrix. Since these effects cannot easily be disentangled experimentally, we use density functional theory (DFT) calculations to estimate their relative contributions to the observed peak broadening. The implantation process is very dynamic and can impart non-equilibrium atomic configurations into the sample. For the current generation of devices, no annealing is used to attempt to heal the lattice after implantation. Here, we consider a variety of plausible atomic environments to explore the range of their effects on capture peak broadening. We focus on $^7$Li because the relaxation of the core hole occurs after the EC decay of $^7$Be to $^7$Li. In section 3.1, the Li 1$s$ binding energies at substitutional and interstitial sites of the bcc Ta lattice are calculated as reference cases for K capture. Section 3.2 then estimates the influence of structural disorder at grain boundaries and locally amorphous sites on the Li 1$s$ energies, and section 3.3 extends the simulations to impurities that are commonly found in sputtered Ta films. Section 3.4 discusses the influence of these effects on the Li 2$s$ levels involved in L-capture events. Finally, we summarize the consequences of the results for high-accuracy BSM physics experiments based on the electron capture decay of $^7$Be to $^7$Li in superconducting sensors, and how this work may indicate that such effects must be evaluated for experiments of this type.

## 2. Computational Details

To study the effect of local atomic structure on the core-hole energies of $^7$Li dopants in Ta, we use Kohn-Sham density functional theory (DFT), which is a computationally efficient method to investigate the electronic structure of solids [7, 8]. In DFT, the many-particle wavefunction is replaced by a set of single-particle orbitals in a transformation that allows the Hamiltonian of the system to be represented as a functional of the electron density. With this transformation, the kinetic energy of the electrons is approximated by that of a non-interacting system, the electrostatic interactions between electrons are approximated by the Coulomb interactions between electron densities, and an exchange-correlation term is used to account for the remaining electronic energy not included in the kinetic energy and electrostatic interactions. The electron-ion interactions, which connect the electron density to an atomistic description of a solid, normally enter the formulation as an external potential. Over the past several decades, DFT has been extensively used to analyze the electronic structure of a wide variety of materials and predict their physical properties associated with details of their atomic arrangements, resulting in tens of thousands of publications.

Our DFT calculations use the projector augmented-wave (PAW) method as implemented in the Vienna Ab-initio Simulation Package (VASP) [9 – 11], with the Perdew–Burke-Ernzerhof (PBE) exchange-correlation functional [12], pseudopotentials containing 13 and 1 valence electrons for Ta and Li, respectively, and spin polarization included. The crystalline system is obtained from a $5 \times 5 \times 5$ cubic supercell of the bcc structure with 250 Ta atoms and with lattice vectors parallel to the [100], [010] and [001] directions. Wavefunctions were expanded in a planewave basis with a 450 eV energy cutoff. Electronic structure calculations were performed using a $2 \times 2 \times 2$ $k$-point mesh (Γ-centered) with Methfessel-Paxton smearing of width 0.10 eV for Brillouin zone integrations. The core-hole energy analyses of Li in Ta are based either on a crystalline system with body centered cubic (bcc) structure or on an amorphous system. The amorphous structure is constructed based on this 250-atom system and is described in more detail below. All structural optimizations were performed until all forces were below 0.01 eV/Å. All core-hole energy calculations were performed by setting the ICORELEVEL tag to 2 in VASP. Convergence of total energies and core-hole eigen-energies were performed to 0.1 meV and 1 meV, respectively.

## 3. Results

### 3.1. Li in Crystalline Ta

As a reference case, we initially consider Li dopants in a perfectly crystalline Ta lattice. When deposited on Al, Ta in STJ detectors grows in the bcc phase. Bulk bcc Ta has a lattice constant $a$ = 3.30 Å (we obtain 3.31 Å with DFT), and all Ta atoms have eight nearest neighbors at a distance of $\sqrt{3}/2$ a = 2.86 Å. Interstitial sites at the center of the cube's faces have octahedral symmetry with four in-plane neighbors at a distance of $a/\sqrt{2}$ = 2.33 Å and two out-of-plane neighbors at a/2 = 1.65 Å (Fig. 2a). In addition, there are two interstitial sites with tetraheral symmetry along each of the midsections of the cube's faces. In the two simplest cases, Li implanted into the Ta lattice can substitute for a Ta atom at one of the bcc lattice sites or be located at an octahedral interstitial site (Fig. 2b). Li is not stable in the tetrahedral interstitial sites and spontaneously moves to the octahedral site with no energy barrier. For Li in an octahedral interstitital, the two out-of-plane neighbors are pushed to a nearest neighbor distance of 2.19 Å while the four in-plane neighbors move to a distance of 2.39 Å. With Li on a substitutional site, the Ta lattice relaxes symmetrically around the Li atom and the nearest neigbors are displaced inward by only a small distance of 0.004 Å, even though the atomic radii of Li (1.82 Å), and Ta (2.20 Å) differ considerably.

For each atomic configuration, we calculate the 1$s$ eigenenergies of Li and the associated energy shift relative to the substitutional case, $\Delta_{1s}$, as

$$\Delta_{1s} = E_{1s}^x - E_{1s}^{subst.} + \Delta E_{Fermi} + \Delta V. \tag{1}$$

Here, $E_{1s}^X$ and $E_{1s}^{subst.}$ are the Li 1s core-hole energies of supercells with Li at a site $x$ of interest and with Li at a substitutional site, respectively, $\Delta E_{\text{Fermi}}$ is the shift in the Fermi levels between these calculations, and $\Delta V$ is a correction to appropriately align the electrostatic potentials of the bulk supercell with the electrostatic potential in the bulk-like regions of each supercell containing Li. We choose the Li binding energy at a substitutional site as the reference energy througout this paper. This choice of $E_{1s}^{subst.} = 0$ in equation (1) is arbitrary and could be replaced by another reference energy without change in conclusions. For a single Li ion at an octahedral interstitial site, the 1s binding energy is reduced by 1.18 eV compared to the substitutional Li (Table I). This suggests that core-hole energies are significantly sensitive to local crystal environment and lattice distortions, even for a well-ordered crystal.

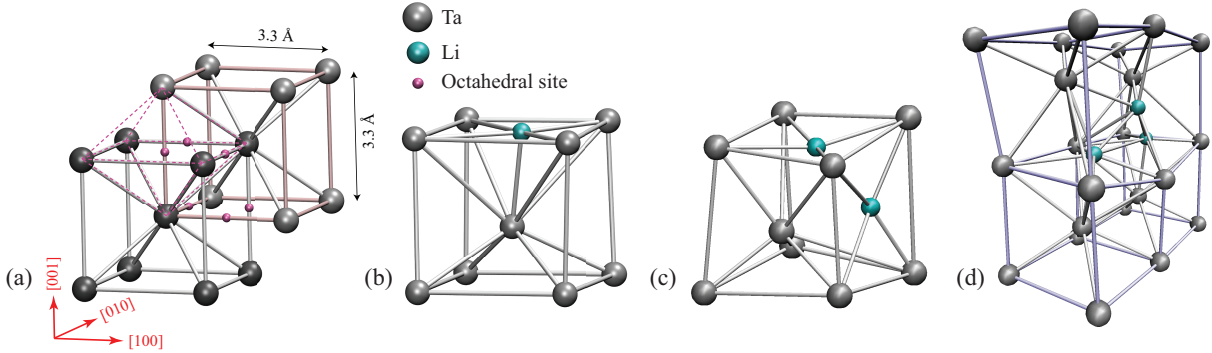

*Figure 2: (a) Bulk Ta has a bcc crystal structure, which is comprised of two intertwined simple cubic lattices where the corner of one cube is the body-center of the other. There are six octahedral interstitial sites per unit cell, which lie at the centers of each face and edge of the cubes; several representative sites are indicated by the small magenta balls. bcc Tantalum with (b) one, (c) two, and (d) three clustered Li interstitials are favorable structures, with the associated local lattice distortions shown. (Only portions of the computational supercells are shown for clarity.)*

To address the question whether additional Li ions (from the large [7]Li background in the [7]Be implantation beam) will settle preferentially near existing Li (Be) implants, we calculate the formation energy $E_f$ of a defect according to

$$E_f = E_{\text{Ta,bulk}+n\text{Li}} - E_{\text{Ta,bulk}} + \mu_{\text{Ta}} - \mu_{\text{Li}}, \quad (2)$$

Here $E_{\text{Ta,bulk}}$ is the total energy of the supercell of 250 Ta atoms in a bcc lattice, $E_{\text{Ta,bulk}+n\text{Li}}$ is the total energy after addition of the Li dopants, and $\mu_{\text{Ta}}$ and $\mu_{\text{Li}}$ are the chemical potentials of Ta and Li, respectively, given by the number of Ta and Li atoms in the supercell and referenced to the pure metals ($\mu_{Ta}^0 = -14.30$ eV/atom and $\mu_{Li}^0 = -1.897$ eV/atom). For our base case of a substitutional Li at a bcc lattice site, the formation energy is $E_f = 1.23$ eV, while at an interstitial site (Fig. 2b) it is 4.07 eV (Table I). We also assess the propensity for interstitial clustering by computing the defect binding energy for multiple adjacent Li interstitials (Table I). We define the defect binding energy as the difference in formation energies for the bound vs. isolated defects, via $E_b = E_{f,\text{bound}} - \sum E_{f,\text{isolated}}$, where negative values indicate preferred binding. We find that two adjacent Li interstitials are favorable, with a total formation energy 3.56 eV higher than the single interstitial and a negative binding energy of –0.26 eV/Li (or –0.51 eV total binding energy). Three clustered Li interstitials are even more energetically favorable compared to three isolated interstitials, with a differential formation energy for the third interstitial of 3.60 eV and binding energy of –0.33 eV/Li (–0.98 eV total binding energy). These results suggests that Li atoms might accumulate in the vicinity of other Li (Be) dopants. In these models, we consider one of the Li as associated with the decay of an implanted Be atom, while the others in the cluster represent background implanted Li. Since the implant process also creates vacancies in the Ta lattice (vacancy formation energy of 2.80 eV), we further considered the case of a Li substitutional dopant bound to an adjacent Ta vacancy, the $Li_{Ta}$–$V_{Ta}$ complex, which is energetically favorable with binding energy of –1.53 eV (Table I). This vacancy complex is associated with a Li $\Delta_{1s}$ shift of 1.67 eV.

| Li site | 1s Energy Shift, $\Delta_{1s}$ [eV] | Formation Energy, $E_f$ [eV] | Defect Binding Energy per Atom, $E_b$ [eV/Li] |
|---|---|---|---|
| Substitutional | 0 | 1.23 | – |
| Single interstitial | 1.18 | 4.07 | – |
| Two interstitials | 1.58 | 7.63 | –0.26 |
| Three interstitials | 1.30 | 11.23 | –0.33 |
| $Li_{Ta}$–$V_{Ta}$ complex | 1.67 | 2.50 | –1.53 |

*Table I: The shift in Li 1s binding energy associated with different lattice incorporation of Li (Be) dopants in crystalline bcc Ta can reach up to 1.67 eV. The 1s binding energy for substitutional Li is used as a reference throughout this paper, defining $\Delta_{1s}$(substitutional) = 0 eV in the first row. The formation energies (and defect binding energies) show that (clusters of) Li interstitials are probable, in addition to substitutional Li. The defect complex in the final row consists of a substitutional Li adjacent to a Ta vacancy.*

Since the binding energies for Li clustering suggest Li may accumulate near other Li (Be) sites and the $^7$Be is accompanied by ~50 times more $^7$Li atoms during implantation, we have also simulated the extreme case where up to the entire first or second coordination spheres around a $^7$Be or $^7$Li dopant are replaced by other $^7$Li atoms. In these cases, the 1s level does shift further, with the simulations showing –1.14 eV ≤ $\Delta_{1s}$ ≤ +0.92 for different configuration of nearest neighbors and a total range of Li 1s binding energies of 2.72 eV.

We further investigated how extreme local strain in the lattice near the implanted atom might further shift the Li 1s binding energy for an otherwise perfect bcc host crystal by displacing the $^7$Li atom along the [100], [110] or [111] lattice directions without relaxing the host atoms. Displacements up to 1 Å lead to local pressure increases up to ~100 GPa, but only yield $\Delta_{1s}$ ~ 0.5 eV. For even larger (unrealistic) displacements approaching half the equilibrium Ta–Ta distance, we obtain $\Delta_{1s}$ less than 2 eV.

Thus, we find that insertion of $^7$Li atoms into the ordered bcc Ta lattice, depending on whether interstitial or substitutional and whether clustered, contributes a measurable and significant shift in the 1s binding energy of up to 1.7 eV, but is not sufficient to account for the entire observed 6.7 eV width of the K-capture peak of the STJ detectors. The maximum broadening of the electron capture peaks occurs when the implanted atoms (Li) are segregated inside the Ta lattice, although we have no evidence so far that such segregation occurs in the STJ detectors.

## 3.2. Li in Disordered Ta

We now consider situations in which the Ta matrix is not perfectly crystalline but shows a certain degree of disorder, which is common in sputtered films and following the high-energy implantation process, especially since annealing is not possible in STJ detectors due to potential damage to the thin tunnel barrier. The most common and obvious structural defect is the presence of grain boundaries in polycrystalline films, but dislocations and local disorder from implant damage tracks may also play a role. Transmission electron microscope (TEM) images of the polycrystalline Ta absorber films in our STJ radiation detectors show average grain sizes of ~20 nm. To understand the change in Li $1s$ core-hole energies at grain boundaries, we consider representative grain boundary structures of the symmetric (210)[001] ("$\Sigma_5(210)$") and (310)[001] ("$\Sigma_3(310)$") tilt grain boundaries in Ta [13], both of which have been extensively studied in the literature [14]. We place $^7$Li atoms in 4 different interstitial and 4 substitutional sites in the vicinity of the grain boundary planes to sample the variety of local environments and allow the structures to fully relax. For both of these grain boundaries, we found that $\Delta_{1s}$ varies between ±0.30 eV for substitutional Li and from –0.30 to +1.65 eV for interstitial Li (Fig. 3).

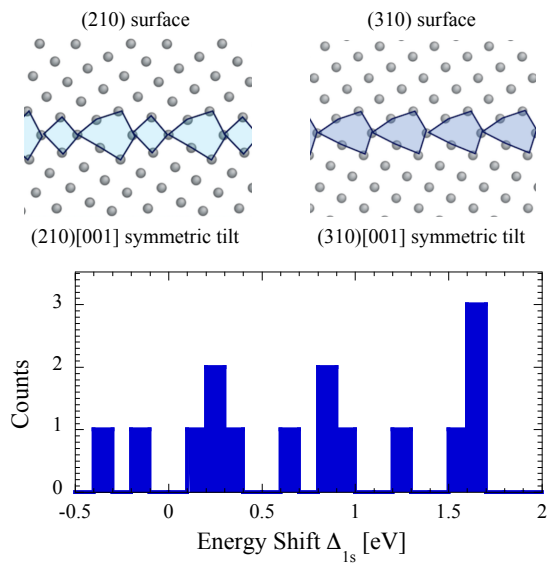

*Figure 3: (Top) Models of grain boundary structures in Ta. (Bottom) The distribution of 1s energy shifts $\Delta_{1s}$ calculated for Li at 4 different substitutional and 4 different interstitial sites within the shaded regions of the grain boundary structures spans –0.3 to 1.65 eV.*

We have also modeled the effect of dislocations on Li $1s$ core-hole energies. In bcc metals such as Ta, low temperature deformation mechanisms are typically governed by the motion of ⟨111⟩/2 screw dislocations, and previous studies have shown that ⟨111⟩/2 dislocations can have different types of core structures [15 – 17]. For our analysis of core-hole energies, we considered two types of dislocation core structures – easy- and hard-core structures – and they were obtained by applying the continuum-scale elastic displacement field of a screw [18]. The two different core structures correspond to either a positive or a negative displacement field along the [111] dislocation line direction. Both of these dislocation core structures preserve the 3-fold rotation crystal symmetry along the [111] direction parallel to the dislocation line and also in the bulk Ta. The easy-core structure corresponds to the ground state configuration, while the hard-core structure has slightly higher energy. When interstitial and substitutional Li atoms were incorporated into the easy-core dislocation structure, we found that the shifts in Li $1s$ core-hole energies were very small, i.e., in the range of $\Delta_{1s}$ = 0.12-0.18 eV. However, when a substitutional

Li was placed in the hard-core structure, we found that $\Delta_{1s} = -0.84$ eV. Similarly, when an interstitial Li was inserted into the hard-core structure, we found that $\Delta_{1s} = -1.62$ eV. These high values can be attributed to small Li–Ta distances in the hard-core structure, as close as 1.77 Å for interstitial Li. The $\Delta_{1s}$ shifts obtained from the dislocation core structures, analyzed as a function of Li–Ta distance, are comparable to the results obtained from our studies of Li displacements along ⟨111⟩ in the crystalline Ta structure.

To further study the extreme limiting case of disorder, we consider the possibility of fully amorphous structures as an approximate model of local implant damage. We note that fully amorphous structures are not realistic for Ta metal, but we aim here to bound the range of these effects. To generate an amorphous Ta distribution, we used a supercell with 250 Ta atoms and performed *ab initio* molecular dynamics melt-quench simulations within the isothermal-isochoric (NPT) ensemble using 1 fs time-steps, a Langevin thermostat with a damping coefficient of 10 ps$^{-1}$, and other parameters as given in Section 2. The system was heated to 6000 K over 10 ps and thermalized for 4 ps, then quenched to 300 K over 2 ps. The atomic positions were then further relaxed by minimizing the energy of the final structure. The resultant structure exhibits a considerable amount of disorder in the distribution of atomic positions (Fig. 4 inset), although the radial distribution function indicates weak short-range order (peaks) up to the second neighbor shell.

To analyze the distribution of core-hole energies for substitutional Li in disordered Ta, we generated 100 structures by randomly replacing one of the Ta atoms with Li and allowing the structure to relax. For interstitial Li in amorphous Ta, we performed a Voronoi tessellation of the amorphous structure and generated 100 structures by placing a Li atom at a randomly selected Voronoi vertex and relaxing them. Figure 4 shows the distributions of Li 1s core-hole shifts obtained from these 200 structures, which range from –0.3 to 1.6 eV. These values are for fully relaxed structures, including volume relaxation. If instead the volumes are constrained to the bulk bcc Ta lattice vectors to approximate local stresses from the implant damage tracks, we observe nearly rigid shifts of these $\Delta_{1s}$ distributions, with the mean $\Delta_{1s}$ increasing by ~0.65 eV from an induced pressure of approximately 9 GPa.

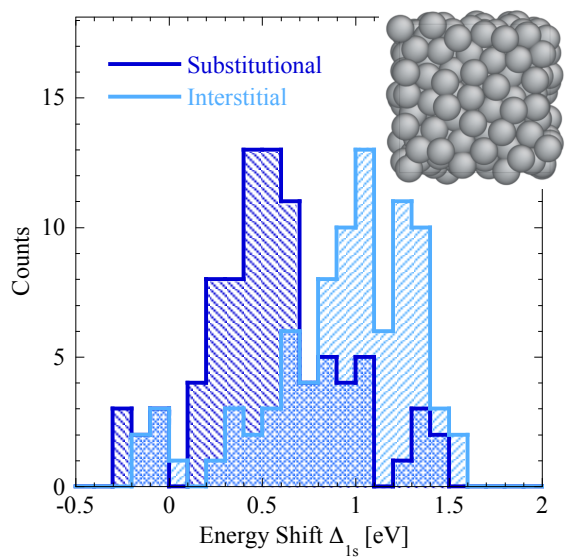

*Figure 4: Distribution of Li 1s energy shifts $\Delta_{1s}$ at 100 substitutional and 100 interstitial sites in amorphous Ta. The inset shows the simulated distribution of Ta atoms after melt-quenching and relaxation.*

Overall, the set of simulations described in this section suggest that locally disordered Ta, which can result from the film deposition process and/or the Be (Li) implantation, results in Li 1$s$ energy shifts $\Delta_{1s}$ between approximately –0.9 and +1.7eV. Thus, similar to site disorder in the crystalline material, structural disorder of the host Ta can explain a significant fraction of the observed broadening of the K-capture peaks, but not the full extent.

### 3.3. Li with Impurities in Ta

Finally, we consider the effect of impurities on the binding energies of the $^7$Be / $^7$Li implants. Various impurities are common in the sputtered films and can originate either from impurities in the initial sputter target or can be incorporated from background impurities in the vacuum chamber during deposition, since Ta is a strong getter material. Impurities can also diffuse into the detector during photolithographic processing or storage, or they can originate from contaminants in the implantation beam of the radioactive $^7$Be dopants, as is the case for the significant background of non-radioactive $^7$Li. We used time-of-flight secondary ion mass spectrometry (ToF-SIMS) to determine the impurities in the Ta absorber film of the STJ detector used to acquire the spectrum in Figure 1. The ToF-SIMS shows that the Ta had a uniform distribution of 1200 ppm of Nb, 500 ppm of W, and 300 ppm of In impurities, which likely originate from the initial Ta sputtering target. The quantifications are based on Ta standards with known doses of implanted elements and have an uncertainty of ±20%. In addition, O and C are present at concentrations of 1000 ppm each, and H and Si at 200 ppm. The concentrations of O, C, H, and Si are highest near the surface, suggesting that they diffused into the Ta film after sputter deposition. Also, O, C and H may have been incorporated during photolithography, and Si may have been sputtered off a Si collimator that was placed in front of the wafer during $^7$Be implantation to define the implantation regions. Additional impurities of F, Na, Cl, Mg, Ca, K, Cr, Mn, Fe, Ni and Cu are detected in single-digit ppm trace quantities.

To evaluate the effects on $\Delta_{1s}$ due to these impurities, we systematically replaced one or more Ta atoms in the first or second coordination spheres around a Li dopant in the crystalline Ta host with one of the elements detected in our Ta films in significant quantities (H, Be, C, N, O, Al, Si, Nb, In). Since there is an almost unlimited number of possible combinations of impurities and configurations, we picked several representative combinations of Li sites and neighboring atoms that are likely to be present in our Ta films. We again further consider some extreme cases to estimate upper limits of the chemical shifts due to impurities. For the extreme cases, we replaced the entire coordination shell around a Li site with selected impurities, i.e., either eight impurities in the first or six in the second shell. The computed shifts of the Li 1$s$ level for the different combinations of impurity atoms and configurations are summarized in Fig. 5, and a few representative structures are depicted in Fig. 6. (All atomic structures are available upon request.) Somewhat surprisingly, the range of shifts $\Delta_{1s}$ is only ~2 eV for most of the likely impurities and configurations, comparable in magnitude to the shifts for pure Li in different sites of the bcc Ta lattice or even at grain boundaries. The effect is larger for the extreme cases with entire coordination spheres replaced with impurity atoms, extending up to ±2 eV, but this situation is not likely to be present in our detectors unless there is strong segregation of impurities or formation of secondary phases. We currently do not have experimental evidence of these occurring, but we cannot completely rule them out. In summary, we find marginal to appreciable effect on Li 1$s$ binding energy in Ta from nearby impurities, but the magnitude of the effects likely does not broaden the EC peaks significantly beyond the levels caused by structural disorder.

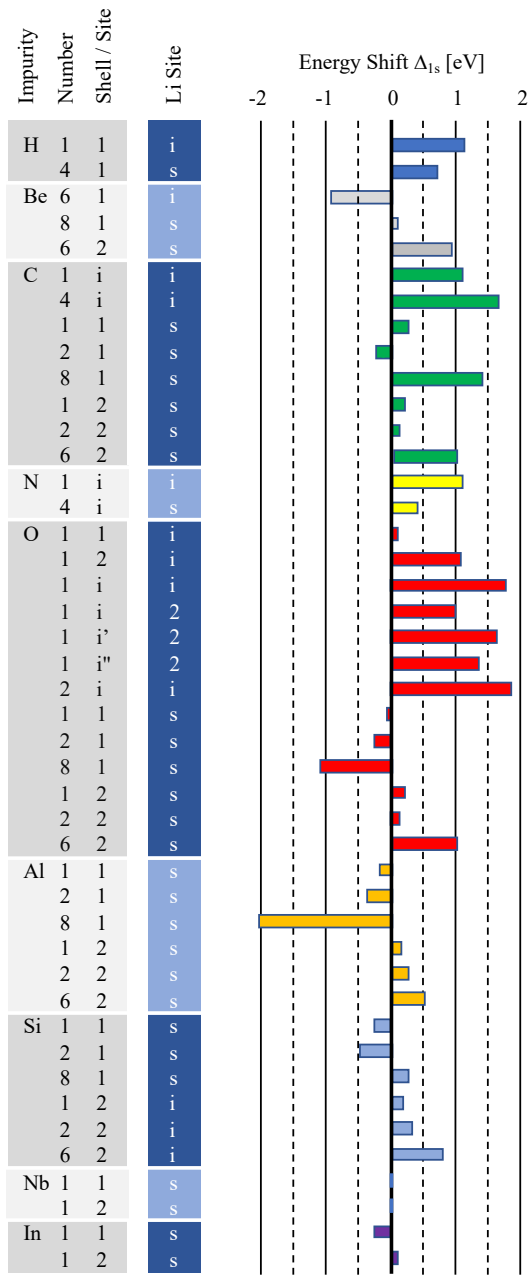

Figure 5: Energy shifts $\Delta_{1s}$ of the Li 1s level due to different impurities near Li dopant(s). The 2nd column ("Number") indicates the number of impurity atoms of the given element considered for each configuration. Li atoms in substitutional sites are denoted by numbers in the 3rd column ("Shell/Site"), which indicate in which neighbor shell the impurities reside, while interstitial sites are denoted by "i." The notation "i" or "s" in the 4th column ("Li Site") indicates whether the adjacent Li dopant sits in an interstitial or substitutional site, respectively, while "2" indicates a special configuration with 2 Li dopants in a dumbbell configuration. The notations i, i' and i" in the 3rd column, for those O impurity configurations with Li dumbbell nearby, refer to different positions of the O impurity in the vicinity of the Li dumbbell. Full neighbor shells consist of 8 or 6 atoms for the first or second neighbor shell, respectively.

One notable set of observations is related to the effects of oxygen, whose influence on the Li 1s energy differs significantly for interstitial and substitutional sites in the bcc Ta lattice (Fig. 6). In a supercell with a single Li interstitial, e.g., Fig. 6(b), the presence of an O interstitial leads to a large shift $\Delta_{1s}$ = 1.76 eV in the Li 1s core-hole energy. However, in a supercell with two Li interstitials in a dumbbell geometry, e.g., Figs. 6(c) and (d), the presence of interstitial O at different positions around the dumbbell causes a shift ranging from 0.97 to 1.62 eV (Fig. 5). Interestingly, when O is located on substitutional Ta lattice sites neighboring the Li, the shifts are very small (< ±0.3 eV). In general, O impurities next to substitutional Li also show small shifts, unless many O atoms (6 or 8) are clustered around the Li. Thus, subtle changes in the position of O impurities around a Li atom can affect the Li 1s core-hole energies significantly. This may reflect the tendency of alkaline atoms such as Li to form strong bonds with O, whose formation depends sensitively on their relative distance. Since O is observed to be one of the highest concentration impurities in our Ta films and shows large variations on the effect on $\Delta_{1s}$ depending on configuration, it could dominate the extent of impurity-related K-capture peak broadening in the Ta STJs; C, H, and N also show notably large shifts without clustering of impurities.

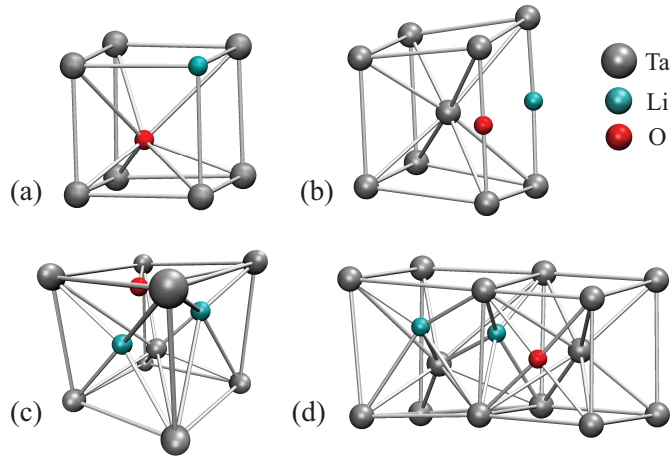

Figure 6: Representative structures containing an O impurity near implanted Li, showing the local lattice distortions. (a) Li and O on neighboring substitutional sites, (O,1,1,s) or row 23 in Figure 5; (b) Li and O on neighboring interstitial sites, (O,1,i,i) or row 18 in Figure 5; (c) Li interstitial dumbbell with neighboring O interstitial, (O,1,i,2) or row 19 in Figure 5; (d) Li interstitial dumbbell with neighboring O interstitial, (O,1,i'',2) or row 21 in Figure 5. (Only portions of the computational supercells are shown for clarity.)

### 3.4. Li 2s levels in Ta

Since the observed broadening of the K-capture and L-capture peaks in the $^7$Be spectra is different (Fig. 1), we also analyzed the effects of site and structural disorder on the Li 2s levels in a Ta matrix. The situation for the Li 2s levels is fundamentally different from the 1s, since the 2s energies overlap with the Ta valence band composed of Ta 5d and 6s states and thus hybridize with the Ta states, creating an inherent broadening. Here, we consider the extent of this broadening and further broadening caused by structural variations of the dopant chemical environment. Figure 7 shows the site-projected electronic density of states (DOS) for Ta, as well as Li at both substitutional and interstitial sites in a crystalline bcc Ta host, in the energy range of the Ta bands near the Fermi level. The Ta DOS we compute is consistent with earlier simulations for pure Ta [19] and does not change significantly for Ta atoms close to a Li defect. The Li 2s levels are hybridized with the Ta levels and spread over most of the Ta band of almost 8 eV. However, the extent of strong hybridization (DOS peak) that affects the STJ L-capture response function depends on whether the Li is incorporated substitutionally or interstitially. Strong hybridization occurs over a significantly broader energy range for interstitial Li (~5 eV) than substitutional Li (~2 eV). Overall, this broadening, especially for interstitial Li, is comparable to the observed width of the L-capture peak, and this electronic hybridization likely causes it to be broader than the K-capture peak, as observed. The Li 2p levels also show similar broadening but are less important in the context of the BeEST sterile neutrino search because their wavefunctions overlap less with the $^7$Be nucleus, so that their contribution to the L-capture signal is negligible.

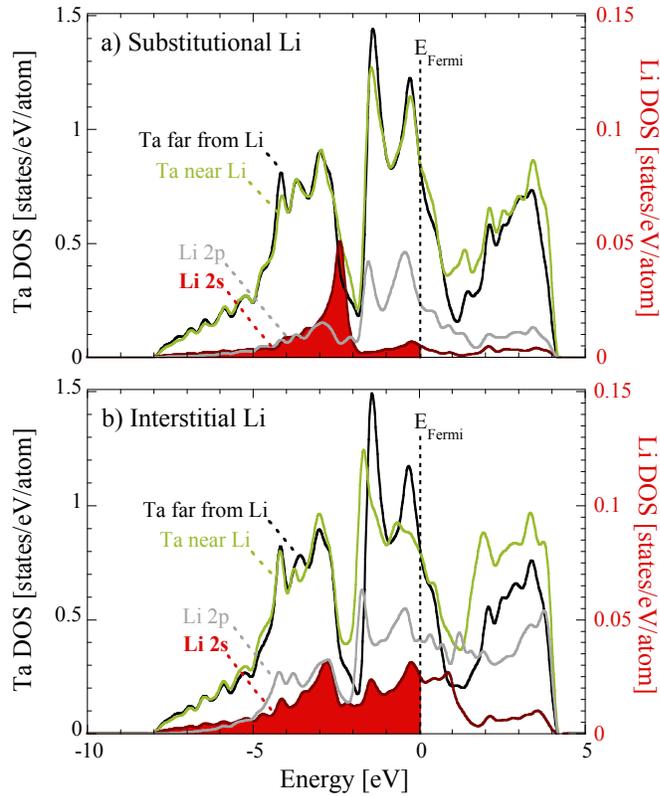

Figure 7: The density of states of Li 2s (red) and 2p (grey) atomic levels for (a) substitutional and (b) interstitial Li in a crystalline bcc Ta host show hybridization with the Ta valence band (black and green lines) and consequent broadening. The red shading indicates the occupied portion of the Li 2s band, which mainly contributes to the $^7$Be L-capture peak.

Finally, we examined how crystalline disorder affects the Li 2$s$ levels, using the same methodology described in Section 3.2 for amorphous Ta structures. Figure 8 shows the corresponding Li 2$s$ DOS for a sampling of 35 random substitutional positions in amorphous Ta. The results indicate additional broadening of the substitutional Li 2$s$ DOS from the atomic disorder, comparable to the extent of hybridization of the interstitial 2$s$ DOS in crystalline Ta. Overall, the effects of hybridization are stronger than the effects of lattice disorder, suggesting the L-capture peak broadening to be fundamentally limited to a larger value than the K-capture for $^7$Be implanted in a metal matrix (e.g., Ta), especially if interstitial implanted dopants are present. The chemical and materials-related effects described above for 1$s$ levels and K-capture, which theoretically could be reduced by atomic-level control of the material, do not provide opportunity to reduce the 2$s$ level (L-capture) broadening beyond the fundamental limit of the orbital hybridization, which will be present to a different extent for any metal with a DOS extending below the Li 2$s$ level (5.4 eV below the Fermi level). This hybridization presents a fundamental lower bound to the L-capture broadening.

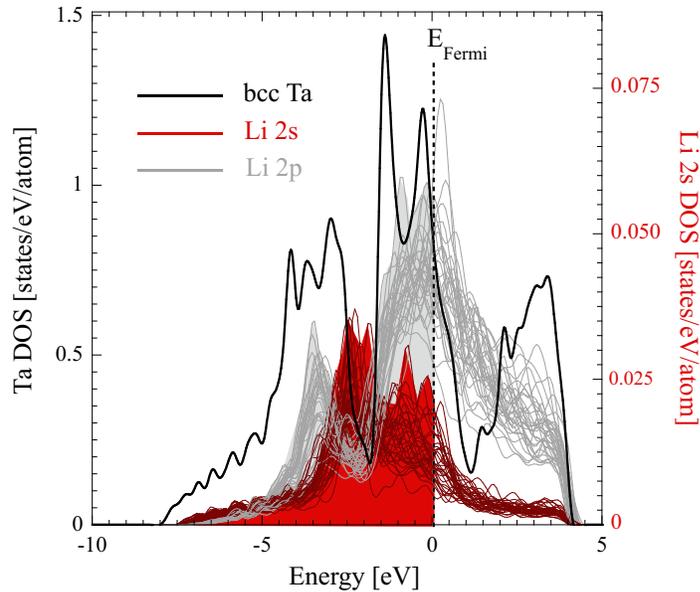

Figure 8: Density of state of 2s (red) and 2p (grey) levels for 35 different substitutional Li sites in amorphous Ta. The shaded states below the Fermi energy are occupied, and only the red 2s orbitals contribute to the electron capture process. Compare to the 2s DOS for substitutional and interstitial Li in crystalline Ta shown in Fig. 7.

## 4. Conclusions

The impact of chemical shifts on the observed EC spectra of implanted radioisotopes due to the host material has been theoretically investigated in detail for the first time. For this, we have performed density functional theory (DFT) simulations of the electronic structure of Li dopants in a polycrystalline Ta matrix relevant to the BeEST experiment. Our work shows that variations in the symmetry and nearest-neighbor configurations of the Li implantation site affect the Li binding energies in a significant way. The energy of the Li 1$s$ level differs from the corresponding value for Li in a substitutional site by approximately –0.5 to +1.7 eV for typical site symmetries and dopant concentrations in sputtered Ta films. Somewhat surprisingly, extreme levels of disorder (amorphization) or very high concentrations of impurities (clusters) only increase the range of Li 1$s$ energy shifts up to a range of ~4 eV. The Li 2$s$ levels are additionally broadened by hybridization with the 5$d$ band of the Ta matrix to a width of over 5 eV. These variations in binding energy contribute to broadening of the electron K- and L-capture peaks from $^7$Be decay inside Ta-based superconducting tunnel junction (STJ) sensors, because the 1$s$ and 2$s$ hole energies of the $^7$Li daughter add to the recoil energy upon relaxation. Such broadening exceeds the STJ detector resolution of 2.0 eV, although the observed broadening (6.7 and 8.2 eV FWHM for the K- and L-capture peaks, respectively) is larger than that predicted from the material disorder effects considered here. This suggests these effects to be only one contribution to the broadening, although they can explain a significant fraction of the overall effect. Furthermore, hybridization of the Li 2$s$ with the Ta 5$d$ orbitals likely causes the L-capture peak to be fundamentally broader than the K-capture peak in the experimental spectra, with little opportunity for sharpening the L-capture peak through extensive material control.

The simulations suggest various approaches to reduce the contribution of material effects to the broadening of the electron capture spectra. One is to start with ultra-pure metals as targets for the absorber film deposition. In addition, electron beam deposition of the STJ electrodes would produce purer films than the current sputtering process. Finally, the absorber film can be annealed after $^7$Be implantation to remove metastable configurations and place the dopants more uniformly into the energetically most stable substitutional site (Table 1). Since STJs cannot be annealed without damaging the tunnel barrier, the dopants must be implanted into the base electrode of the STJ so that the film can be annealed before the formation of the junction. The relatively long $^7$Be half-life makes this approach quite feasible. This would also lessen the impact of diffusion of materials into the STJ during storage because the dopants in the base electrode will be well protected.

This study quantifies how chemical effects can broaden the EC decay spectra in solid-state radioisotope-doped cryogenic detectors. Understanding this broadening mechanism for $^7$Li in Ta thin films, in particular, is significant because it currently limits the search for sterile neutrinos in the mass range below 100 keV/c$^2$ with the BeEST experiment [3]. Our findings suggest that detailed materials imaging of the devices to determine the specific environmental conditions for the embedded atoms, combined with these DFT calculations, may allow identifying the sources of spectral broadening. Further, these theoretical results may also provide a path towards materials engineering of the system in the future to improve the K-capture sensitivity by reducing the in-medium effects and variations in the desired system. Nonetheless, it appears very difficult to limit structural and chemical inhomogeneities to contribute less than order 1 eV to the capture peaks in practical devices. Similar broadening will likely also occur to some extent in other electron capture experiments with radioactive isotopes implanted into cryogenic detectors. Although the relative magnitude of these effects may be smaller in other systems, it is clear that they must be evaluated in detail for experiments of this type to characterize possible systematic uncertainties.

**Acknowledgements**

This work was funded by the LLNL LDRD grant 20-LW-006 and the Office of Nuclear Physics in the U.S. Department of Energy's Office of Science under Grant DE-SC0021245. This work was performed under the auspices of the U.S. Department of Energy by Lawrence Livermore National Laboratory under Contract No. DE-AC52-07NA27344. All structures will be made available upon reasonable requests.